\documentstyle[11pt]{article}

\newcommand{\ba}{\begin{array}}
\newcommand{\ea}{\end{array}}
\newcommand{\be}{\begin{equation}}
\newcommand{\ee}{\end{equation}}
\newcommand{\bea}{\begin{eqnarray}}
\newcommand{\eea}{\end{eqnarray}}
\newcommand{\beas}{\begin{eqnarray*}}
\newcommand{\eeas}{\end{eqnarray*}}
\newcommand{\ra}{\rightarrow}

\font\blackboard=msbm10 at 12pt
\font\blackboards=msbm7
\font\blackboardss=msbm5
\newfam\black
\textfont\black=\blackboard
\scriptfont\black=\blackboards
\scriptscriptfont\black=\blackboardss
\def\bb#1{{\fam\black\relax#1}}

\begin{document}

\pagestyle{plain}
\setcounter{page}{1}

\baselineskip14pt

\begin{flushright}
SNU-TP-98-098\\
hep-th/9808166\\
\end{flushright}
\vspace{.5cm}

\begin{center}
{\Large More on Membranes in Matrix Theory}\\
\vspace{.5cm}
{\large Nakwoo Kim 
\footnote{Address as of October 1st, 1998: Department of Physics,
Queen Mary and Westfield College, Mile End Road, London E1 4NS UK; 
{\tt N.Kim@qmw.ac.uk}}
}\\
\vspace{0.5cm}
{\it Department of Physics and Center for Theoretical Physics,}
\\ {\it  Seoul National University, Seoul 151-742, Korea} \\
{\tt nakwoo@phya.snu.ac.kr}
\end{center}
\vskip 0.5 cm
\begin{abstract}
We study noncompact and static membrane solutions 
in Matrix theory. Demanding axial symmetry on a membrane
embedded in three spatial dimensions,
we obtain a wormhole solution whose shape is the same with
the catenoidal solution of Born-Infeld theory. We also discuss
another interesting class of solutions,  
membranes embedded holomorphically in four spatial dimensions,
which are $1/4$ BPS.
\end{abstract}
\setcounter{page}{1}
\renewcommand{\thefootnote}{\arabic{footnote}}
\setcounter{footnote}{0}

Matrix theory \cite{bfss}, formulated from the $U(N)$ gauge supersymmetric 
quantum mechanics describing motions of $N$ D0-branes, is now well 
established as a nonperturbative formulation of M-theory. The basic properties
of M-theory which could be readily checked were that it becomes 11 
dimensional supergravity at low energy, and it contains extended objects
like membrane and its magnetic dual five-brane. 
And the dynamics of D-particles of Type IIA superstring theory 
converges to the supermembrane action in the lightcone frame,
formulated in \cite{membrane}.
In addition 
to the rather trivial flat branes, compact branes, in particular 
with spherical topologies 
were obtained and studied also \cite{taylor,rey,D4M}. 
These branes with finite size are not static, but oscillating in time.
Recently an interesting class of static membrane solutions was presented in 
\cite{HoloCurve}, viz. holomorphically embedded membranes. 
In this note we investigate the possibility of another static 
solution of Matrix theory. We find one embedded in three dimensions
which describes wormhole membrane, made by a string between branes
and antibranes.
Similar system was already discussed using Born-Infeld (BI) theory 
\cite{CalMal,Gibbons}, which
is low energy effective theory for Dp-brane. It is amusing to see that our 
solution has the same shape with the electrostatic 
solution of three dimensional Born-Infeld theory, 
describing two D2-branes connected by a throat.

In this note it is sufficient to consider only the bosonic sector of 
the Matrix theory lagrangian,
\be
L = {\rm Tr} \left( \frac{1}{2g} (D_t X^I)^2 + \frac{1}{4g} 
( [ X^I , X^J ]^2 ) \right),
\label{mlag}
\ee
where $X^I,I=1,\ldots,9$ are $N\times N$ hermitian matrices and 
$g$ is the string coupling constant. $D_t X \equiv \partial_t X - i [A,X]$
where $A=X^0$ is the gauge field of our gauged quantum mechanics.
It is well known that Matrix theory, in the large $N$ limit, converges
to the 11 dimensional supermembrane theory in lightcone frame.
In other words, Eq.(\ref{mlag}) becomes
\be
L = \frac{p_{11}}{2} \int dp dq ( D_t X^I )^2 
+ \frac{1}{4p_{11}} \int dp dq ( \{ X^I, X^J \} )^2
\ee
viz. the trace is approximated by integration over two 
variables and the matrix commutator by $i$ times Poisson bracket,
\be
\{ X(p,q) , Y(p,q) \} \equiv i \left(
\frac{\partial X}{\partial q} \frac{\partial Y}{\partial p} -
\frac{\partial X}{\partial p} \frac{\partial Y}{\partial q}
\right).
\ee
$q,p$ are the worldsurface coordinates of the membrane.
The equation of motion for above lagrangian is
\be
D_t^2 X^I + [X^J,[X^J,X^I]] = 0,
\ee
or in large $N$ limit,
\be 
D_t^2 X^i + \{X^j,\{X^j,X^i \} \} = 0.
\ee
For simplicity we set the coupling constants to 1 from now on.

The simplest solution is flat membrane
\bea
X^1 &=& \sqrt{\frac{q}{2}} \cos p, \nonumber \\
X^2 &=& \sqrt{\frac{q}{2}} \sin p, \nonumber \\
X^J &=& 0 \;\;\;\;\; J \neq 1,2, \label{flat}
\eea
where $p$ is an angular variable and $q$ can take any real number.
It is obvious that above parametrization expands the entire range of
$12$-plane. For Matrix theory we take ladder operators $a,a^\dag$ 
of harmonic oscillator and set $X^1=(a+a^\dag)/\sqrt{2},
X^2=i(a-a^\dag)/\sqrt{2}$. 
This basic membrane solution is $1/2$ BPS of Matrix theory.

Another interesting example is that of spherical shape,
\bea
X^1 &=& r(t) \sqrt{1-q^2} \cos p, \nonumber \\
X^2 &=& r(t) \sqrt{1-q^2} \sin p, \nonumber \\
X^3 &=& r(t) q, \nonumber \\
X^J &=& 0 \;\;\;\;\; J \neq 1,2,3.
\eea
We have $S^2$ with radius $r(t)$, which isn't a constant but oscillates
according to the equation $\ddot{r} + 2 r^3 = 0$. Note that in this case 
we can obtain finite dimensional representation of the solution, i.e.
$X^i = r(t) J_i \;\; (i=1,2,3)$, where $J_i$ are the familiar 
angular momentum operators. But for the infinite flat membranes
we can't satisfy the equation of motion in terms of matrices
with finite size.

Lately Cornalba and Taylor studied the problem of finding
static solutions in Matrix theory \cite{HoloCurve}. They
studied solutions which can be represented by holomorphic
curves, i.e. membranes embedded holomorphically in ${\bb C}^4$.
To define holomorphically embedded membranes we introduce 4
complex coordinates, 
\bea
Z_1 &=& X^1 + i X^2, \nonumber \\
Z_2 &=& X^3 + i X^4, \nonumber \\
Z_3 &=& X^5 + i X^6, \nonumber \\
Z_4 &=& X^7 + i X^8, 
\eea
and truncate the last spatial coordinate $X^9 = 0$. By a holomorphic 
curve we mean they are all holomorphic functions of one complex variable, 
$Z_A = f_A(z)$. 
After quantization $z$ is traded into an operator or matrix.
In terms of above complex variables the potential term of the membrane theory
can be written as
\be
V = -\frac{1}{16} \sum_{A,B=1..4} \int dq dp \left(
\{ Z_A , Z_B \} \{ \bar{Z}_A, \bar{Z}_B \} 
+
\{ Z_A , \bar{Z}_B \} \{ \bar{Z}_A, Z_B \} 
\right).
\ee
Assuming $\{z,\bar{z}\}=F(z,\bar{z})$, minimizing the potential we have
\be
F(z,\bar{z}) \left( \sum_A f'_A(z) f'_A(\bar{z}) \right) = C,
\ee
where $C$ is a constant. Note that this condition means
\be
\sum_{A} \{ Z_A , \bar{Z}_A \} = 
\sum_{A} \{ f_A(z) , f_A(\bar{z}) \} = 
C. \label{bps}
\ee
This was introduced as a {\it gauge condition} in \cite{HoloCurve}, but
as we have seen from above this can be obtained as a result of 
the equation of motion for the membrane theory, and we can show that
even in Matrix theory, where $Z_A$ become noncommutative, this condition
makes the equation of motion trivially satisfied. So we admit the 
following is BPS condition for a static membrane in Matrix theory
\be
\sum_A [Z_A , \bar{Z}_A ] = C \label{bpsM}.
\ee
In fact this condition guarantees preservation of some proportion of
supersymmetry. Consider the supersymmetry transformation of fermion fields
in Matrix theory, 
\be
\delta \theta = \frac{1}{2} \Gamma_I D_t X^J \epsilon + 
\frac{i}{4} \Gamma_{IJ} [X^I,X^J] \epsilon + \epsilon'  ,
\label{sutra}
\ee
where $\Gamma_{IJ}=\frac{1}{2} [\Gamma_I,\Gamma_J]$, and $\Gamma_I,
I=1,\ldots,9$ are nine-dimensional gamma matrices which satisfy
$\Gamma_I\Gamma_J + \Gamma_J\Gamma_I = \delta_{IJ} {\bf 1}_{16\times16}$.
There are two 16-component supersymmetry parameters $\epsilon, \epsilon'$,
the former is ordinary dynamic supersymmetry parameter, while the latter
is called kinematic one, which is common in lightcone formulation. Since here
we are interested in static BPS solutions without nontrivial gauge 
field background, the first term vanishes.
Assume we have membrane solution holomorphically embedded in $\bb C^2$,
viz. $Z = X^1 +i X^2,W = X^3 + i X^4$ and 
\be
[Z,W] = [\bar{Z},\bar{W}] = 0.
\ee
And expand the second term of Eq.(\ref{sutra})
\bea
\frac{1}{4} \Gamma_{IJ} [X^I,X^J] \epsilon &=&
\frac{i}{4} \left( \Gamma_{12} [Z,\bar{Z}] + \Gamma_{34} [W,\bar{W}]  
\right) \epsilon \nonumber \\
&& + \frac{1}{8} ( \Gamma_{13} + \Gamma_{24} ) 
\left( [Z,\bar{W}]+[\bar{Z},W] \right) \epsilon 
\nonumber \\
&& +\frac{i}{8} (\Gamma_{14} - \Gamma_{23} )
\left( [Z,\bar{W}] - [\bar{Z},W] \right) \epsilon   .
\eea
Now if we have 
\be
\Gamma_1 \Gamma_2 \Gamma_3 \Gamma_4 \epsilon  = - \epsilon,
\label{hoone}
\ee
the second and third term vanish. Now applying the BPS condition 
Eq.(\ref{bpsM})
\be
[Z,\bar{Z}] + [W, \bar{W}] = C,
\ee
we obtain additional condition for supersymmetric solution,
\be
-\frac{1}{4} C \Gamma_{12} \epsilon + \epsilon' = 0 .
\label{hotwo}
\ee
With Eq.(\ref{hoone}) and Eq.(\ref{hotwo}), we see 
$1/4$ of the supersymmetry is conserved by holomorphic
membrane solution of Matrix theory when it is embedded 
in $\bb C^2$.

As the simplest example we can consider membranes intersecting at
right angle, represented by $Z=z,W=1/z$.
We take ansatz $z = f(q) e^{ip}$ and $\{ q,p \}=i$. Then it is straightforward
that Eq.(\ref{bps}) means
\be
\left( f^2 - \frac{1}{f^2} \right)' = C. \label{intermem}
\ee
Solving it we have the following solution for a static membrane
\bea
X^1 &=& \sqrt{\frac{Cq + \sqrt{C^2 q^2 +4}}{2}} \cos p, \nonumber \\
X^2 &=& \sqrt{\frac{Cq + \sqrt{C^2 q^2 +4}}{2}} \sin p, \nonumber \\
X^3 &=& \sqrt{\frac{- Cq + \sqrt{C^2 q^2 +4}}{2}} \cos p, \nonumber \\
X^4 &=& \sqrt{\frac{- Cq + \sqrt{C^2 q^2 +4}}{2}} \sin p, \nonumber \\
X^I &=& 0 \;\;\;\; I \neq 1,2,3,4.
\eea
As $q \ra \infty$ the membrane lies on $12$-plane, while 
in the limit $q \ra -\infty$ it lies on $34$-plane. 
Thus this solution represents membranes intersecting at a point, 
and the fact that this is $1/4$-BPS is consistent with the 
general result of \cite{IntMat}, which studied supersymmetric
intersecting brane configurations of Matrix theory.

Because
of the ordering problem the above solution in itself does not turn out to
be useful in obtaining 
the solution of Matrix theory directly. Instead we are hinted from the
flat membrane solution Eq.(\ref{flat}) and its interpretation
in terms of harmonic oscillator operators. We set
\bea
X^1 &=& \frac{1}{2} ( \delta_{i,j+1} x_j + \delta_{i,j-1} x_i ) 
, \nonumber \\
X^2 &=& \frac{i}{2} ( \delta_{i,j+1} x_j - \delta_{i,j-1} x_i ) ,
\eea
and $X^3,X^4$ according to the curve equation $W=1/Z$. 
Now we solve the equation of motion from the Matrix theory
lagrangian to obtain the following difference equation,
\be
x^2_{i+1} - \frac{1}{x^2_{i+1}} =
x^2_{i} - \frac{1}{x^2_{i}}  + C,
\ee
which is obviously the {\it quantized} version of the membrane 
equation Eq.(\ref{intermem}).
With a constant $C$ and initial value of $x_i$ we can 
calculate every term of the array recursively. Calculating 
the eigenvalues of the coordinates, we note that 
\bea
(X^1)^2 + (X^2)^2 &=& 
\frac{1}{2} {\rm Diag} (\ldots,x^2_{i+1}+x^2_i,\ldots) ,
\nonumber \\
(X^3)^2 + (X^4)^2 &=& 
\frac{1}{2} {\rm Diag} (\ldots,x^{-2}_{i+1}+x^{-2}_i,\ldots) .
\eea
When $C$ is positive $x^2_i \ra \infty$ as $i \ra \infty$,
and $x^2_i \ra 0$ as $i \ra -\infty$, so the D0-branes
with label $i$ very large are confined near the origin of
34-plane but very far from the origin of 12-plane, and 
oppositely when $i$ goes to negative infinity. This again 
leads us to interprete the solution as intersecting membranes. 

Now we turn to finding a membrane embedded nonholomorphically in 
three spatial dimensions. As ansatz we demand axial symmetry,
and this time we try a solution with nontrivial excitation of the
gauge field. Since Matrix theory describes M-theory in lightcone
frame, the time component of the gauge field corresponds to the
excitation of the membrane in $X^-$ direction. 
This should be exactly the way to get a fundamental string from
M-theory in DLCQ (discrete lightcone quantization) formalism. 
We'll find that our solution corresponds to the BI solution of
fundamental string attached to a D-brane.
\bea
X^0 &=& h(q), \nonumber \\
X^1 &=& f(q) \cos p, \nonumber \\
X^2 &=& f(q) \sin p, \nonumber \\
X^3 &=& g(q), \nonumber \\
X^I &=& 0 \;\;\;\; I \neq 0,1,2,3. \label{bion}
\eea
Using the equation of motion the functions $f,g,h$ should satisfy
\bea
\frac{1}{2} (f^2)'' &=& (g')^2 - (h')^2, \nonumber \\
( f^2 g' )' &=& ( f^2 h' )' \; = \; 0,  \label{worm}
\eea
where prime denotes differentiation with respect to the variable $q$.
Being autonomous,
this set of coupled nonlinear differential equations is easily 
integrated. We have
\bea
\label{shape}
g &=& \pm \int \frac{C df}{\sqrt{k^2 f^2 - (C^2 - D^2)}} \nonumber \\
&=& \pm \frac{C}{k} \log \left( f + \sqrt{f^2 -(C^2-D^2)/k^2} \right), \\
\label{efield}
h &=& \pm \int \frac{D df}{\sqrt{k^2 f^2 - (C^2 - D^2)}} \nonumber \\
&=& \pm \frac{D}{k} \log \left( f + \sqrt{f^2 -(C^2-D^2)/k^2} \right), \\
q &=& \int \frac{2 f^2 df}{\sqrt{k^2 f^2 - (C^2 - D^2)}}
\nonumber \\
&=& 
\frac{1}{k} \left(
f \sqrt{f^2 - (C^2-D^2)/k^2} - \frac{C^2-D^2}{k^2} \log ( f +
\sqrt{f^2 - (C^2-D^2)/k^2}  
\right),
\eea
where $C,D,k$ are integration constants.

Let's have a look at
the supersymmetry transformation rule Eq.(\ref{sutra}), under
the ansatz Eq.(\ref{bion}).
\bea
\delta \theta &=& 
- \frac{1}{2} f h' ( \sin p \Gamma_1 - \cos p \Gamma_2 ) \epsilon
-\frac{1}{2} f f' \Gamma_{12} \epsilon \nonumber \\
& & -\frac{1}{2} f g' \sin p \Gamma_{13} \epsilon
+\frac{1}{2} f g' \cos p \Gamma_{23} \epsilon
+ \epsilon' .
\eea
When $C=D$, the solution simplifies into
\bea
f &=& \sqrt{kq} , \nonumber \\
g &=& h \; = \; \pm \frac{C}{2k} \log q.
\eea
and it is straightforward to see that when $C=D$ the solution is
$1/4$ BPS with the following condition of unbroken supersymmetry.
\bea
( 1 + \Gamma_3 ) \epsilon &=& 0 , \nonumber \\
-\frac{k}{4} \Gamma_{12} \epsilon + \epsilon' &=& 0.
\eea

M-theory membrane is dual to D2-brane of Type IIA string theory.
Since Born-Infeld theory describes low energy effective dynamics of D-branes,
we expect that both theories may allow the same solutions. 
\footnote{This fact was noted also in \cite{BorHop},
that the three dimensional Born-Infeld action can be derived from the action
of relativistic membrane moving in ${\bb R}^3$ through gauge fixing.}
In \cite{CalMal,Gibbons} solutions with transverse
excitation with electromagnetic charge were found,
\bea
\label{catshape}
X(r) &=& \int^\infty_r \frac{B}{\sqrt{r^{2p-2} - r^{2p-2}_0 }} dr,
\\
\label{eshape}
E &=& F_{0r} = \frac{A}{\sqrt{r^{2p-2} - r^{2p-2}_0}},
\eea
where X represents one of the transverse direction of the p-brane, and
$E$ is the radial component of the electric field. $r$ is the radial
coordinate on the worldvolume,
$r_0^{2p-2} = B^2 - A^2$, and the BPS condition corresponds to $A/B \ra 1$,
or $r_0 \ra 0$. When $r_0=0$ this solution represents a string attached
on D-branes, while $r_0 \neq 0$ we have catenoidal solution of brane
antibrane bound state.
We find the form of Eq.(\ref{catshape}),(\ref{eshape})
is the same with Eq.(\ref{shape}),(\ref{efield}), 
and furthermore in the same limit of point charge solution, 
the two solutions become BPS, as expected.

For the matrix representation we follow the same reasoning
which was used for the holomorphic curve $W=1/Z$, and set 
\bea
X^1 &=& \delta_{i,j+1} x_j + \delta_{i,j-1} x_i, \nonumber \\
X^2 &=& i ( \delta_{i,j+1} x_j - \delta_{i,j-1} x_i ), \nonumber \\
X^3 &=& \delta_{i,j} y_i,  \nonumber \\
X^0 &=& \delta_{i,j} z_i,  \nonumber \\
X^I &=& 0 \;\;\;\;\; I \neq 0,1,2,3.
\eea
And try to solve the equation of motion for static configuration.
We have the following coupled differential equations for $x_i,y_i,z_i$.  
\bea
x_{i+1}^2 - 2 x_i^2 + x_{i-1}^2  &=& 
\frac{1}{2} ( y_{i+1} - y_i )^2 - 
\frac{1}{2} ( z_{i+1} - z_i )^2  ,
\nonumber \\
x^2_i ( y_{i+1} - y_i ) &=& x^2_{i-1} ( y_i - y_{i-1} ), \nonumber \\
x^2_i ( z_{i+1} - z_i ) &=& x^2_{i-1} ( z_i - z_{i-1} ).
\eea
It is evident that this is just the discretized version of the 
coupled differential equations Eq.(\ref{worm}), so  
it is reasonable to admit the solution
of above difference equations as quantized version of the 
continuum solution. Above equations can be rewritten as
\bea
x^2_{i+1} &=& 2 x^2_i - x^2_{i-1} + 2 \frac{C^2 - D^2}{x^4_i}, \nonumber \\
y_{i+1} &=& y_i + \frac{C}{x^2_i}, \nonumber \\
z_{i+1} &=& z_i + \frac{D}{x^2_i}.
\eea
This matrix solution also becomes BPS when $C=D$.

It would be excellent if we repeat the study of dynamic issues 
treated in \cite{CalMal} with Matrix theory solution and find
coincidence with supergravity again. But unfortunately the BI theory 
result is claimed to be incompatible with the supergravity calculation except 
for D3 and D4 branes \cite{LPT}.

One interesting topic for further study is the extension to 
higher dimensional branes. In \cite{D4M} 
4 dimensional spherical branes in Matrix theory were constructed
using $SO(5)$ gamma matrices. It would be
exciting if we could obtain 4 dimensional static solutions which have
the same shape with (4+1)D Born-Infeld theory solution. 
Since in that case the result of 
BI fluctuation study is identical to the supergravity result,
we might do the same calculation in terms of Matrix theory and check
if it is consistent with BI or supergravity. 

\end{document}